\documentclass{article}

\usepackage{PRIMEarxiv}

\usepackage[utf8]{inputenc} 
\usepackage[T1]{fontenc}    
\usepackage{hyperref}       
\usepackage{url}            
\usepackage{booktabs}       
\usepackage{amsfonts}       
\usepackage{nicefrac}       
\usepackage{microtype}      
\usepackage{lipsum}
\usepackage{authblk}
\usepackage{fancyhdr}       
\usepackage{graphicx}       
\graphicspath{{media/}}     

\title{jaxsgp4: GPU-accelerated mega-constellation propagation with batch parallelism
}

\author[1,2,3]{Charlotte Priestley}
\author[1,2,3]{Will Handley}
\affil[1]{Kavli Institute for Cosmology, University of Cambridge, Cambridge, UK}
\affil[2]{Institute of Astronomy, University of Cambridge, Cambridge, UK}
\affil[3]{PolyChord Ltd., London, UK}

\begin{document}
\maketitle

\begin{abstract}
As the population of anthropogenic space objects transitions from sparse clusters to mega-constellations exceeding 100,000 satellites, traditional orbital propagation techniques face a critical bottleneck. Standard CPU-bound implementations of the Simplified General Perturbations 4 (SGP4) algorithm are less well suited to handle the requisite scale of collision avoidance and Space Situational Awareness (SSA) tasks. This paper introduces \texttt{jaxsgp4}, an open-source high-performance reimplementation of SGP4 utilising the \texttt{JAX} library. \texttt{JAX} has gained traction in the landscape of computational research, offering an easy mechanism for Just-In-Time (JIT) compilation, automatic vectorisation and automatic optimisation of code for CPU, GPU and TPU hardware modalities. By refactoring the algorithm into a pure functional paradigm, we leverage these transformations to execute massively parallel propagations on modern GPUs. We demonstrate that \texttt{jaxsgp4} can propagate the entire Starlink constellation (9,341 satellites) each to 1,000 future time steps in under 4 ms on a single A100 GPU, representing a speedup of $1500\times$ over traditional C++ baselines. Furthermore, we argue that the use of 32-bit precision for SGP4 propagation tasks offers a principled trade-off, sacrificing negligible precision loss for a substantial gain in throughput on hardware accelerators.
\end{abstract}


\section{Introduction}
The orbital environment is undergoing unprecedented expansion. As of March 2026, despite only 12,200 active satellites currently in orbit\footnote{\url{https://www.space-track.org/}, data retrieved 25th March 2026},  the Federal Communications Commission (FCC) and International Telecommunication Union (ITU) have received licence applications for 1,896,629 planned constellation satellites to be launched in the coming decades\footnote{\url{https://planet4589.org/space/con/conlist.html}, data retrieved 25th March 2026}. This trajectory is only accelerating \cite{falle2023}; for instance, SpaceX’s proposal for `orbital data centres' in low Earth orbit (LEO) added an additional one million satellites to the global forecast in January 2026. While not all proposed assets will reach orbit, the immediate need to future-proof our current infrastructure has already been recognised. The U.S. Space Force `Space Fence' system, operational since 2020, is expected to track an order of magnitude more than the 26,000 objects managed by its predecessor. Similarly, the industry-standard Two-Line Element (TLE) set format is being phased out due to its inability to support catalogue numbers above 99,999.

In this era of mega-constellations, the ability to perform massively parallel calculations will be vital for Space Situational Awareness (SSA), which underpins the safety and continuity of satellite operations. Such capacity is equally critical for mitigating the contamination of our telescopes amid mounting concerns over the impacts of mega-constellations on the future of astronomical observation \cite{borlaff2025}. 

Orbital propagation (the task of predicting the future position and velocity of a satellite from a known state), is the computational kernel of these applications. The Simplified General Perturbations 4 (SGP4) model from the simplified perturbations family (SGP, SGP4, SDP4, SGP8, SDP8) \cite{hoots1980} remains the most ubiquitous propagator. This popularity reflects the fact that the TLE data format (the de-facto universal standard for distributing publicly available satellite data) is specific to this family of models, such that anyone working with TLE data must employ a simplified perturbations model to get accurate predictions. For ease of adaptability, SGP4 also has the benefit of being open source and the underlying equations easily accessible. The history and development of the theory as well as the current equations can be found in \cite{hoots2004}.

The source code introduced in \cite{vallado2006}, available in full at the CelesTrak website\footnote{\url{https://celestrak.org/publications/AIAA/2006-6753/}}, is widely-recognised as the most comprehensive and up-to-date version of SGP4 available to the public. While this `official' release is available in several languages, including C++, FORTRAN, MATLAB, and Pascal, these implementations are designed for serial execution or limited multi-threading. They are less well-suited to vectorise across the millions of satellite-time combinations required by modern conjunction assessment screenings.

Modern probabilistic programming languages such as \texttt{JAX}\footnote{\url{https://jax.readthedocs.io}} and \texttt{PyTorch} \cite{paszke2019} offer an attractive solution to this computational bottleneck. \texttt{JAX} provides a frontend for XLA (Accelerated Linear Algebra), allowing Python code to be compiled and optimised for hardware accelerators (GPUs and TPUs) via a NumPy-like interface. It has seen wide adoption in scientific and numerical computing owing to its simple mechanisms for Just-In-Time (JIT) compilation, automatic vectorisation, and automatic differentiation. We emphasise that these probabilistic programming frameworks provide two distinct capabilities: automatic differentiation and GPU-accelerated batch parallelism. While the two are often conflated, they serve fundamentally different purposes; automatic differentiation is essential for gradient-based optimisation, whereas parallelism leverages the huge processing power of GPUs to enable 1000s of tasks to be processed simultaneously. In practice, batch parallelism is often significantly underappreciated. This is likely because most practitioners' intuitions about GPUs are shaped by training deep learning models, where one selects the largest network that will fit on a single GPU, or even distributes it across many, leaving no room to run multiple instances in parallel. In scientific computing, however, models are orders of magnitude smaller than trillion-parameter neural networks, meaning that running a single evaluation on a GPU leaves the vast majority of the device idle. This represents a largely untapped axis of computational power: by batching many independent evaluations, one can exploit a factor of thousands more parallelism than a single evaluation alone.

SGP4 has already been implemented in one such framework: $\partial$SGP4\footnote{\url{https://github.com/esa/dSGP4}}, built on \texttt{PyTorch}. However, this effort focused on the benefits of a differentiable implementation for gradient-based analysis, treating the potential for massive parallelism as a secondary convenience~\cite{acciarini2025}. As a result, $\partial$SGP4's memory scaling is not optimised for massively-parallel workloads (see Section~\ref{performancebenchmark}). Yet, it is this parallelism that may prove the more impactful capability: recent work has shown across scientific applications that exploiting batch parallelism on GPUs can actually be a much more powerful axis than gradients, including for example, in applications to cosmology~\cite{lovick2025}\cite{leeney2025}\cite{ormondroyd2025} and gravitational wave inference~\cite{prathaban2025}. Orbital propagation is a natural fit for this paradigm as each satellite's trajectory can be computed independently. Moreover, unlike `special perturbation' techniques, which numerically integrate the equations of motion such that each timestep depends sequentially on the last, SGP4 is a `general perturbation' technique, providing an analytical solution to the equations of motion such that the state at any time can be evaluated directly from the initial conditions. SGP4 therefore yields two independent potential axes for parallelism, over satellites and over times, making it especially well-placed to reap the benefits of GPU acceleration. It is this untapped potential that motivates us to explore the prospects of GPU-accelerated batch parallelism as a solution to the increased computational demands of an exponentially growing satellite population.

In this work, we present \texttt{jaxsgp4}, a \texttt{JAX}-native implementation of SGP4 for massively-parallel propagations that can be implemented on modern hardware (CPU/TPU/GPU), and make it available through an open-source repository\footnote{\url{https://github.com/cmpriestley/jaxsgp4}}. We demonstrate how refactoring SGP4 for data-parallel execution on hardware accelerators addresses the computational bottleneck of the mega-constellation era. In Section \ref{Implementation}, we give an overview of how the SGP4 propagation model was translated into \texttt{JAX}. In Section \ref{performancebenchmark} we benchmark the performance of \texttt{jaxsgp4} on GPU, and demonstrate its utility for large-scale propagation tasks with the Starlink constellation. We consider the viability of 32-bit propagations in Section~\ref{Precision}, and in Section~\ref{autodiff} we outline \texttt{jaxsgp4}'s automatic differentiation capabilities. We discuss these methods in Section \ref{Discussion} and conclude in Section \ref{Conclusion}.

\section{jaxsgp4: Implementation}\label{Implementation}

The core contribution of this work is the translation of the procedural, state-heavy SGP4 algorithm into a functional architecture compatible with \texttt{JAX}'s transformation system.

\subsection{Architecture and Refactoring}

Standard SGP4 implementations take a state-heavy object-oriented approach, relying on a mutable `Satellite' state to store changing satellite data over time. Additionally, standard SGP4 depends on complex branching logic (e.g., modifying drag modelling depending on an object's perigee altitude). \texttt{JAX}, however, requires `pure' functions---functions that depend only on inputs and have no side effects---to apply its transformations effectively. In this paradigm, a function will always return the same results given the same inputs; functions have no hidden states and can not mutate objects or global variables. 

We rewrote the SGP4 logic into Python directly from the current equations laid out in~\cite{hoots2004} originally taken from~\cite{hoots1980}, with reference to the sgp4 Python library\footnote{\url{https://github.com/brandon-rhodes/python-sgp4}} for implementation specifics. Our implementation uses the standard WGS72 geopotential constants. The initialisation logic (the computation of numerous constant terms that takes place before the main time-dependent propagation algorithm can proceed), often separated in standard libraries, was integrated directly into the \texttt{jaxsgp4} computational graph. This allows the accelerator to handle the full pipeline from TLE parsing to Cartesian state vector output ($\textbf{r}, \textbf{v}$).

To validate this refactoring, we benchmarked the accuracy of \texttt{jaxsgp4} against the standard C++ SGP4 implementation using the active Starlink catalogue. Our tests confirmed that \texttt{jaxsgp4} matches the C++ baseline to within expected machine precision tolerances, including edge cases like near-circular orbits and low-perigee trajectories. Finally, because the codebase relies entirely on \texttt{JAX} array primitives, it is hardware-agnostic (running on CPUs, GPUs, or TPUs without modification) and natively supports both 32-bit and 64-bit floating-point precision. The benefits of this precision flexibility are discussed further in Section~\ref{Precision}.

\subsection{JAX Key Concepts}

Refactoring the algorithm into a purely functional form makes it compatible with the full suite of \texttt{JAX} transformations, including \texttt{jax.grad} for automatic differentiation (see Section~\ref{autodiff}). We leverage three key \texttt{JAX} concepts to achieve the performance gains presented in the main body of this work (Section~\ref{performancebenchmark}): Just-In-Time (JIT) compilation, functional control flow, and automatic vectorisation. 

\paragraph{JIT Compilation} 
To maximise throughput, the entire propagation routine can be Just-In-Time (JIT) compiled using \texttt{jax.jit}. JIT compilation traces the sequence of operations and fuses them into optimized XLA kernels targeted for the specific hardware. For an algorithm like SGP4, which involves hundreds of sequential arithmetic operations, this fusion improves memory efficiency by keeping intermediate variables in fast, on-chip registers. Once compiled, the routine bypasses the Python interpreter entirely, allowing execution speeds comparable or faster than natively compiled C++ code.

\paragraph{Control Flow} 
XLA compilation requires static computational graphs, meaning the path through a function cannot depend on the values of its inputs. However, SGP4 contains significant branching logic, altering calculations based on the values of certain orbital parameters such as epoch perigee height. To make the function compatible with \texttt{JAX}, standard Python \texttt{if/else} statements were replaced with \texttt{JAX} control flow primitives like \texttt{jax.lax.cond}. These allow the compiler to trace both branches and execute them efficiently on parallel hardware.

Along the same lines, traditional implementations include runtime validity checks that further complicate SGP4's branching logic, such as exiting the propagation routine if eccentricity exceeds 1.0 or mean motion becomes negative. As is standard when porting to GPU, we removed these runtime checks from the core propagation routine to conserve functional purity. Instead, anomalous physical states are allowed to compute mathematically and are flagged via error codes, effectively deferring validity filtering to a computationally inexpensive post-processing step.

\paragraph{Vectorisation} 
A key advantage of this implementation is that we may apply \texttt{JAX}'s \texttt{jax.vmap} (vectorising map). Standard SGP4 implementations propagate multiple satellites to multiple time steps using nested, sequential loops. Using \texttt{jax.vmap}, we are able to automatically transform our base propagation function, $f(s,t)\to (\textbf{r},\textbf{v})$
, to operate over arrays. By composing a \texttt{vmap} over the time inputs with a second \texttt{vmap} over the satellite parameters, we can evaluate large batches simultaneously. This approach avoids manual thread management and maps the independent calculations directly to the thousands of cores available on modern GPUs, enabling the simultaneous propagation of entire constellations.

\section{Performance Benchmarking}\label{performancebenchmark}

To assess to what extent exploiting modern hardware accelerators (GPUs) can improve the scalability of satellite propagation, we evaluated \texttt{jaxsgp4} on both NVIDIA T4 and A100 GPUs against the official C++ code found in \cite{vallado2006} on a standard CPU.

\subsection{Experimental Setup}\label{experimental_setup}

Performance benchmarks were conducted in a cloud-based environment (Google Colab). The proposed \texttt{JAX} implementation was executed on both an NVIDIA Tesla T4 GPU with 16 GB of GDDR6 VRAM, and an NVIDIA A100-SXM4 GPU with 40GB of HBM2 memory. The T4 is one of NVIDIA's more cost-effective entry-level models, and while newer-generation GPUs exist, the A100 is still widely used as an industry-standard high-performance card. Traditional SGP4 implementations are inherently single-threaded and CPU-bound; they cannot natively exploit modern hardware accelerators (GPUs). The baseline C++ implementation was consequently executed on an Intel Xeon CPU @ 2.20 GHz. 

The baseline C++ implementation refers to the official code from \cite{vallado2006}, regarded as the most accurate and up-to-date public version of the SGP4 theory since the United States Department of Defence (DoD) first released the equations and source code in \cite{hoots1980}. This was compiled with the \texttt{sgp4} Python package version 2.25\footnote{\url{https://pypi.org/project/sgp4/}} utilising 64-bit precision. The \texttt{JAX} implementation utilised 32-bit precision to better optimise GPU efficiency, this choice is discussed in more detail in Section~\ref{Precision}.

All TLE data used in the analyses of this section was taken from Celes-Trak\footnote{\url{https://celestrak.com}}, and comprises of TLEs for the Starlink constellation with an epoch date of the 13\textsuperscript{th} January 2026. 

\paragraph{Timing Specifics} 
For statistical stability, for each timing measurement in our benchmark an adaptive number of iterations were recorded such that the total execution time for each measurement exceeded 0.2 seconds. Five of such trials were conducted, from which we report the corresponding minimum time taken for a single run of the code. As is standard practice in computer science, the minimum value is reported (rather than the mean or median) on the premise that artefacts caused by system noise, such as transient operating system interrupts, will only slow the process down, and therefore the minimum value is most likely to reflect the true ability of the hardware~\cite{chen2016}.

To ensure a fair comparison of computational throughput, for the \texttt{JAX} implementation the latency associated with transferring data from the CPU host to the GPU device was excluded from timing measurements (by pre-converting inputs to jax arrays), isolating the propagation kernel performance. In practice, this reflects how the code would be deployed: TLE data is loaded onto the GPU once and reused across many propagation calls, meaning the one-off transfer cost is amortised over the lifetime of the application. The measured \texttt{JAX} time also excludes the initial compilation cost (warm-up), representing the steady-state performance expected in a production environment.

Finally, we note that in the C++ case the recorded time only includes the time taken for the main propagation routine; it does not include the time taken for the initialisation phase of the algorithm. This initialisation phase involves the computation of constant terms from the raw TLE data. These constant terms are then used in the propagation routine in combination with the input time to make predictions of satellite motion. Unlike the standard C++ library, \texttt{jaxsgp4} integrates both initialisation and propagation into a single call. As a result, the \texttt{JAX} timings reflect the time taken for the full pipeline from TLE parsing to state vector output $(\textbf{r},\textbf{v})$, and as such give a conservative estimate of \texttt{JAX}'s capability. The propagation phase is the most computationally demanding so we do not expect this to have a major impact on the results. 

\subsection{Scaling Analysis}\label{scaling}

\begin{figure}
    \includegraphics{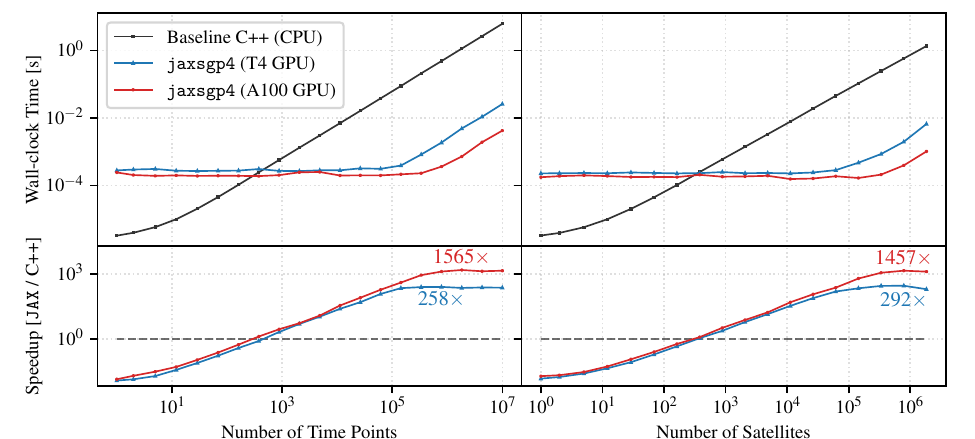}
    \caption{Scaling performance comparison between \texttt{JAX}/GPU and C++/CPU implementations of the SGP4 algorithm, for a single satellite propagated to multiple times (left) and multiple satellites propagated to a single time (right). \texttt{jaxsgp4} exploits the potential of modern GPUs to perform massively-parallel computations, exhibiting a flat scaling regime where increased workload does not increase wall-clock time until hardware saturation (top). The `break-even' point for each GPU (Speedup=1, denoted by dashed line in bottom panels), at which the benefit of parallel computation for large batch sizes overcomes the initial dispatch overhead, occurs at batch sizes of $\sim$300-500. In the bottom panels, we also specify the maximum speedup achieved by \texttt{jaxsgp4} over the C++ baseline. This occurs in the linear hardware saturation regime, at which point the GPUs are operating at maximum computational efficiency.}
    \label{fig:both_plots}
\end{figure}

The main advantage of a refactored SGP4 explored in this paper is that it allows us to exploit the ability of GPUs to perform massively-parallel computations. Having rewritten the algorithm into a pure, functional paradigm compatible with \texttt{JAX}'s transformations, we make it possible to run huge numbers of computations in parallel by applying the \texttt{jax.vmap} (vectorising map) transformation to automatically batch over function inputs. 

This capacity for parallel computations means the time taken to propagate with \texttt{jaxsgp4} does not increase with increased workload until hardware saturation. However, the advantage of a flat scaling regime only becomes apparent at sufficiently large batch sizes; each call to the GPU carries a fixed, one-off cost, which for smaller batches, dominates the total execution time. Below this threshold, codes run directly on CPU with lower overhead will often be faster. Once the batch size is large enough, this fixed dispatch cost is amortised and the GPU processes additional work in parallel at no extra time cost, making it significantly faster than a CPU which must handle each computation sequentially. This break-even point, much like the point of hardware saturation, must be assessed empirically as it depends on the complex interplay between a number of factors such as the specific task considered. 

In light of this, we set out to assess how the time taken for \texttt{jaxsgp4} to complete a propagation task on GPU scales as we batch over each input of the SGP4 propagation routine (increasing the number of time points being propagated towards or the number of satellites to be propagated). The results of this analysis are given in Figure \ref{fig:both_plots} in relation to the standard C++ baseline performance, where we have utilised 32-bit and 64-bit precision for \texttt{jaxsgp4} and C++ respectively in view of the conclusions reached by the cost-benefit analysis in section \ref{Precision}. Computation times are given for propagating up to 10 million time points and $\sim$1.8 million satellites, a workload sufficiently large to explore the hardware saturation regime in each case. Since this required more satellites than currently exist, an artificial catalogue of one million satellites was created by tiling the existing Starlink TLE data (9341 satellites). This ensured the propagation task remained representative of real-world scenarios, while still stressing the GPU to complete each computation in full, no different to the scenario in which all the satellites are distinct. The number of satellites is capped at 1,832,980 as at this point the T4 GPU runs out of memory. It is noteworthy that even low-grade hardware is capable of handling a number of satellites comparable to the total number proposed for future launch across all current mega-constellation filings.

The time taken by the traditional CPU-bound C++ implementation scales linearly with the scale of the task as it can only execute each propagation in sequence. In contrast, \texttt{jaxsgp4} exhibits a flat scaling curve for batch sizes below $10^5$; propagating a single satellite with \texttt{jaxsgp4} on an A100 GPU takes the same wall-clock time as propagating 100,000 satellites thanks to the device's ability to handle the workload in parallel. The advantages of \texttt{jaxsgp4} become apparent at moderate batch sizes, to the order of hundreds of satellites or times, at which point the benefits of parallel computation start to rapidly outweigh the cost of increased computational overhead. The maximum speedup achieved by \texttt{jaxsgp4} over the C++ baseline occurs once the GPU reaches peak utilisation. This is evidenced on the curve by the transition from flat to a linear regime, at which point the GPU reaches its full capacity for parallelism and remains operating at its peak computational efficiency. Using an entry-level T4 NVIDIA GPU, we find \texttt{jaxsgp4} delivers a maximum speedup of around 250$\times$ over the standard C++ routine, when batch sizes exceed of order 100,000 times or satellites. With the more costly, high-performance A100 GPU, \texttt{jaxsgp4} continues to deliver improvements over C++ past this scale, until it is as much as 1500$\times$ faster than the standard routine for propagation tasks involving batches of order $10^6$. 

\subsection{Full-Catalogue Propagation}

\begin{figure}
    \includegraphics{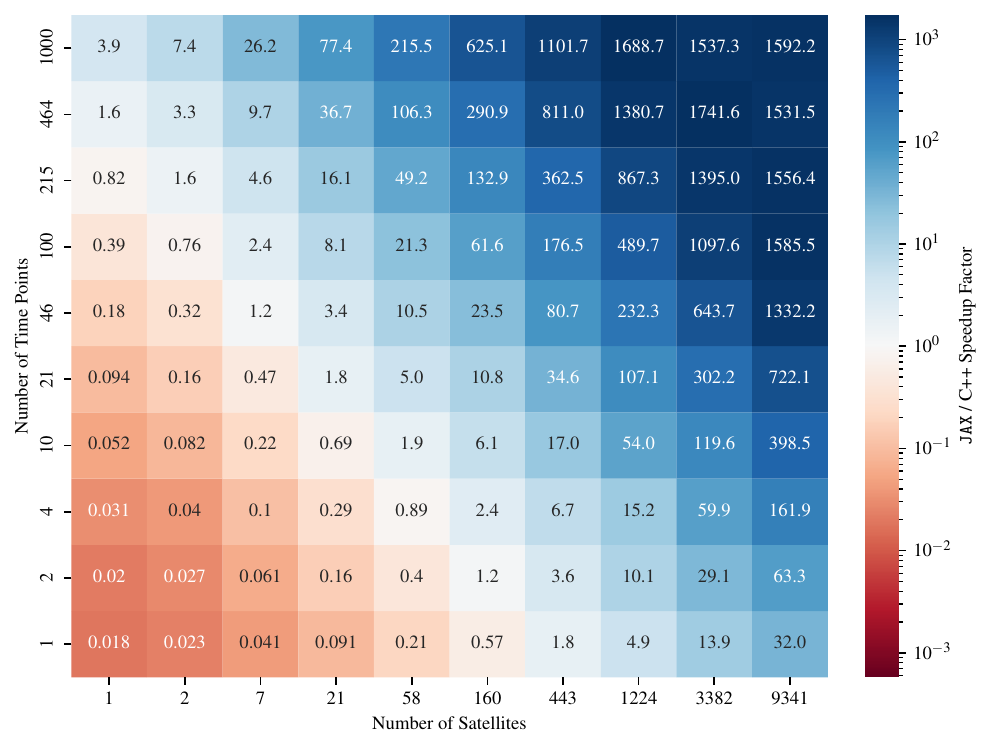}
    \caption{GPU-accelerated \texttt{JAX} vs C++ SGP4 performance comparison.  Blue indicates \texttt{JAX} is faster, Red indicates C++ is faster. The standard C++ implementation is inherently single-threaded and was run on a standard CPU while the \texttt{JAX} benchmark was run on an NVIDIA A100 GPU. Log-spaced axes show the relative performance of each implementation across different scales.}
    \label{fig:heatmap}
\end{figure}

Following a discussion of scaling specifics, we now consider a more realistic propagation scenario to demonstrate the advantages of GPU-accelerated \texttt{jaxsgp4} for large-scale SSA tasks. Figure \ref{fig:heatmap} shows the speedup achieved by \texttt{jaxsgp4} over traditional CPU-bound C++ routines for a range of scenarios involving propagating a number satellites each to a number of future times. The \texttt{jaxsgp4} implementation was run on an NVIDIA A100 GPU utilising 32-bit precision. 

We observe two distinct performance regimes. For tasks involving below order $10^2$ individual propagations, such as taking 21 satellites each to 10 future times, the traditional serial-bound C++ implementation actually outperforms \texttt{jaxsgp4} at light computational loads. The true advantage of GPU-accelerated \texttt{jaxsgp4} comes in its ability to outperform traditional propagation routines by orders of magnitude at the large scale, for example, workloads involving simultaneous propagations of an entire constellation of satellites. We find for the full Starlink catalogue propagated to 1,000 time steps, \texttt{jaxsgp4} completes the task on an NVIDIA A100 GPU in 3.8 ms. This represents a speedup of over 1592$\times$ compared to the serial C++ implementation. For a more detailed discussion on scaling specifics, refer to section~\ref{scaling}. 

\section{Precision: 32-bit vs. 64-bit}\label{Precision}

\begin{figure}
    \includegraphics{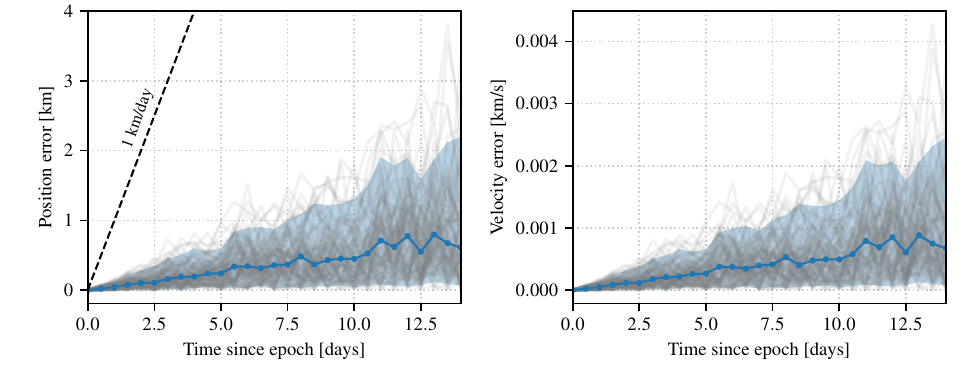}
    \caption{Relative error accumulation over two weeks running \texttt{jaxsgp4} with 32 bit precision versus the standard SGP4 implementation which uses 64 bit precision. The median and 5\textsuperscript{th} - 95\textsuperscript{th} percentile regions are highlighted in blue. The dashed line depicts a conservative lower bound on physical error growth due to limitations of the SGP4 model itself. \texttt{jaxsgp4} errors are order 1e-9, 1e-12 for position and velocity respectively (floating point) when the same analysis is performed using 64 bit precision.  }
    \label{fig:error}
\end{figure}

While SGP4 is traditionally run in 64-bit precision to minimise numerical round-off error accumulation, doing so limits computational throughput on hardware accelerators. Modern GPU architectures are heavily optimised for 32-bit operations (FP32), offering considerably higher Floating Point Operations Per Second (FLOPS) and memory bandwidth utilisation with FP32 over FP64. In this section, we analyse the accuracy trade-off of utilising FP32 for SGP4 propagation tasks. We argue that the resulting floating-point quantisation errors are acceptable when contextualised within the kilometre-scale intrinsic physical inaccuracies associated with the TLE format and SGP4's modelling limitations. 

Predictions made with SGP4+TLEs fundamentally involve errors of several kilometres due to the inaccuracies of the input TLE format, which comprises mean rather than osculating elements, as well as the physical assumptions made by the SGP4 model, for example, in its highly simplified treatment of atmospheric drag. While the TLE format is being superseded by the Orbit Mean-Elements Message (OMM), this is a change in data encoding rather than in the underlying element representation, and the same physical limitations apply. An often-quoted heuristic for SGP4+TLE prediction accuracy is an error of approximately 1km at epoch which degrades by an additional 1-3km a day \cite{levit2011, vallado2006, kelso2007, dong2010}.

To isolate the error introduced purely by reduced numerical precision, we evaluated the output of \texttt{jaxsgp4} running in FP32 against the standard FP64 C++ implementation. The TLE data and implementation specifics used in this analysis are as described in Section \ref{experimental_setup}. To ensure the validity of our baseline, we first verified that an FP64 configuration of \texttt{jaxsgp4} successfully reproduces the results of the standard C++ implementation to within expected FP64 machine epsilon tolerances, specifically yielding deviations on the order of $10^{-9}$ km (micrometres) for position and $10^{-12}$ km/s for velocity.

Figure \ref{fig:error} illustrates the growth of the absolute deviation in predicted position and velocity when propagating 100 Starlink TLEs for a two-week duration using 32-bit \texttt{jaxsgp4}, compared against the standard 64-bit C++ predictions. Typical LEO radius is approximately $6.6 \times 10^3$ metres, and FP32 provides approximately 7 decimal digits of precision. Consequently, the minimum positional quantisation error is roughly 0.7 metres. Over a two-week propagation, we observe that the median positional error associated with precision loss begins at approximately 1 m at epoch and remains under 1 km over the course of the propagation. Even in the 95\textsuperscript{th} percentile of worst-case scenarios, the deviation grows at roughly 2 km per week. When compared to the inherent physical error of the SGP4 model—depicted as a conservative lower-bound growth of 1 km per day (dashed line in Figure \ref{fig:error})—the numerical error introduced by FP32 is dwarfed by the limitations of the propagation routine itself. Velocity deviations exhibit a similarly negligible fractional error, at most on the order of a few metres per second after two weeks. We therefore conclude that 32-bit precision offers a virtuous trade-off, sacrificing negligible operational accuracy for substantial performance gains on modern hardware accelerators. 

\section{Differentiable Orbital Mechanics}\label{autodiff}

Beyond raw computational speed, refactoring SGP4 into a pure functional paradigm in \texttt{JAX} introduces the capability of automatic differentiation (autodiff) to the model. Autodiff allows for fast and exact computation of gradients of the final state vector with respect to the initial orbital elements (e.g., inclination, eccentricity, or the drag term $B^*$). This differentiability allows users to perform gradient-based optimisation of the model and facilitates the integration of SGP4 with modern machine-learning techniques.

The benefits of a differentiable SGP4 implementation have been explored by recent work: Acciarini et al.~\cite{acciarini2025} introduced the PyTorch-based $\partial$SGP4, and proposed its utility for a number of applications including gradient-based orbit determination, exact state transition matrix computation and covariance propagation. They further demonstrate how differentiability enables neural networks to be composed directly with the propagator, improving accuracy beyond baseline SGP4. More recently, Naylor et al.~\cite{naylor2025} have proposed the use of a differentiable propagator to optimise inspection trajectories in order to improve the quality of captured images.

\texttt{jaxsgp4} inherits these capabilities directly; the propagator is composed entirely of differentiable \texttt{JAX} primitives so transformations such as \texttt{jax.grad} and \texttt{jax.jacobian} can be easily applied. Moreover, the composability of \texttt{JAX} transformations means that batched gradients over large catalogues of satellites are obtained simply by combining \texttt{jax.grad} with \texttt{jax.vmap}, requiring no additional implementation effort while benefiting from the same hardware acceleration described in Section~\ref{autodiff}.

We note that $\partial$SGP4, the most established differentiable SGP4 implementation, also supports GPU-accelerated batch propagation. However, our local benchmarks find that \texttt{jaxsgp4} offers a $>10\times$ improvement in speed over $\partial$SGP4. Additionally its $\mathcal{O}(N+M)$ memory scaling compared to $\partial$SGP4's $\mathcal{O}(NM)$ enables \texttt{jaxsgp4} to handle large workloads that exceed $\partial$SGP4's GPU memory capacity.

\section{Discussion}\label{Discussion}

The results presented in this work demonstrate that the transition from serial, CPU-bound propagation routines to parallel, accelerator-native architectures represents more than a mere software optimization; it is a qualitative shift in the scale of computationally feasible astrodynamics. Modern SSA tasks involve enormous computational loads, for example, the continuous evaluation of hundreds of millions of satellite-debris pairs in all-vs-all conjunction screening. In this large-scale regime, we evidence the potential for orders of magnitude improvement over current baselines, where tasks which previously took hours can now be completed in seconds. We expect these findings to become increasingly important in the coming decades, as the rise of mega-constellations and improvements to our tracking capabilities stand to dramatically increase the number of catalogued objects in orbit. \\

Acknowledging this progress, we now consider certain limitations of \texttt{jaxsgp4}. The current \texttt{jaxsgp4} implementation focuses on near-Earth orbits (orbital periods under 225 minutes), as in the original SGP4 formulation first developed by Ken Cranford in 1970 \cite{lane1979}. The Deep Space extensions (SDP4) \cite{hujsak1979}, later integrated into SGP4, are currently under development and require additional, state-dependent branching logic which may complicate XLA compilation. 

Lastly, a small caveat when operating in FP32 with the current \texttt{jaxsgp4} is the representation of the TLE epoch inside the \texttt{jaxsgp4} Satellite object, where it is encoded as a day-of-year plus fractional day (e.g., 316.22557474). Stored as a single FP32 value, the integer day component consumes most of the significant digits. This results in a systematic 'zero error' of approximately 1 second at worst, which at LEO velocities, translates to kilometre-scale position errors. This can be easily resolved in future by updating the Satellite object to store the epoch as separate integer day and fractional day fields. For now, the preferred interface when operating in FP32 is to supply time since epoch in minutes (rather than Julian date) to \texttt{jaxsgp4}'s propagation routine, as this bypasses epoch reconstruction entirely: typical propagation windows of up to 14 days ($\approx$20{,}160 minutes) are well within FP32's representable range, and a value of zero at epoch is exact. \\

Beyond the implementation-specific limitations of \texttt{jaxsgp4}, it is important to acknowledge the broader cost trade-offs of the GPU-accelerated paradigm. While GPUs provide substantial performance improvements, these gains incur significantly higher hardware and operational costs. For example, at on-demand Google Cloud Compute Engine rates\footnote{\url{https://cloud.google.com/compute/all-pricing}, accessed March 2026}, the A100 GPU instance (a2-highgpu-1g) at \$3.67 per hour is approximately 10$\times$ more expensive per hour than a comparable CPU-only instance (n1-standard-8) at \$0.38. However, despite the A100 being more expensive on a per-hour basis, the $\sim$1500$\times$ speedup achieved by the GPU implementation means the total cost to complete a large-scale propagation task is approximately 150$\times$ cheaper on the GPU. Even the more affordable T4 GPU, at roughly 2$\times$ the hourly cost of a comparable CPU instance, achieves a $\sim$250$\times$ speedup, making it substantially cheaper overall for large workloads.

It is important to recognise, however, that these peak performance gains are realised only at large computational scales due to the characteristic scaling behaviour of GPUs. For smaller-scale propagation tasks, the GPU overhead reduces the effective speedup, and the cost advantage diminishes accordingly. The break-even point for cost efficiency over the CPU occurs for tasks involving of order $10^3$ individual computations; for instance, propagating 3000 satellites to a single time costs roughly the same dollar amount on both the CPU and the A100 (though the GPU completes the task approximately 10$\times$ faster). Beyond this threshold, the GPU becomes both faster and cheaper per propagation.

Taken together, these results demonstrate that for large-scale propagation tasks, the speedup achieved by \texttt{jaxsgp4} far outweighs the increased per-hour cost of GPU hardware, making such tasks not only faster but also more cost-effective. Nonetheless, practitioners should consider the scale of their workload when evaluating whether the GPU-accelerated approach offers a meaningful cost benefit over traditional CPU-bound methods. \\

A final aspect worth discussing is the expected performance scaling of \texttt{jaxsgp4} for different GPU models. As observed in Section \ref{performancebenchmark}, the performance characteristics of \texttt{jaxsgp4} are highly dependent on the underlying hardware; orbital propagation is generally a memory-bound operation, therefore performance is generally correlated with GPU memory bandwidth. The NVIDIA A100 utilised in our benchmarks features approximately five times the memory bandwidth of the NVIDIA T4, which aligns with the proportionally higher peak achieved speedup observed in Figure \ref{fig:both_plots}. Memory bandwidth has roughly doubled with each generation of NVIDIA GPUs, leaving a substantial gap between the A100 and successive Hopper (2022) and Blackwell (2025) generations. Relative to the A100, the current frontier (NVIDIA B200) has approximately five times the memory bandwidth, which, assuming the computation remains memory bound, would offer a $\sim7500\times$ speedup over current C++ SGP4 baselines at its peak efficiency. 

\section{Conclusions}\label{Conclusion}

As the orbital population scales exponentially, our computational tools must evolve from serial, CPU-bound paradigms to parallel, accelerator-native architectures. In this paper, we introduce \texttt{jaxsgp4}, a high-performance, purely functional reimplementation of the SGP4 algorithm utilising the \texttt{JAX} framework. We demonstrate the ability of \texttt{jaxsgp4} to dramatically improve the scalability of satellite propagation by exploiting the parallelisation capabilities of modern GPU hardware. Our analysis confirms that refactoring our current propagation algorithms to run on modern GPU hardware does not just represent an optimisation of current code, but a shift in the scale of what is possible, speeding up constellation-wide propagation times by three orders of magnitude over C++ baselines. 

In addition, we argue that the use of 32-bit precision for SGP4 propagations is well-justified, offering a substantial gain in throughput on GPUs whilst introducing only negligible positional errors in the context of the kilometre-scale physical inaccuracies of the SGP4 model itself. Finally, we note that \texttt{jaxsgp4} has the additional benefit of being fully differentiable, facilitating gradient-based optimisation and composability with modern machine-learning techniques.

Future work will focus on integrating SDP4 Deep Space perturbations into the computational graph. Additionally, while SGP4 remains ubiquitous due to its speed and compatibility with the TLE catalogue, its low physical accuracy limits its utility for precision manoeuvrers. Applying the \texttt{JAX} transformation paradigm to high-fidelity semi-analytical models (such as SGP4-XP or Gauss's Variational Equations) or high-precision numerical propagators, could yield tools capable of matching SGP4's traditional runtime speeds while delivering the accuracy required for the next generation of space traffic management.

Follow-up work may also explore a range of science applications, in particular, those with large-scale propagation as a central computational requirement. Monte Carlo simulations of the Kessler syndrome, in which cascading collisions are modelled across thousands of stochastic realisations of the full orbital population, are a natural candidate for GPU-accelerated propagation. Similarly, forecasting the contamination of astronomical observations by mega-constellations requires propagating entire catalogues across dense grids of future times and telescope pointings. Beyond batch parallelism alone, gradient-based constellation optimisation (for example minimising collision risk or maximising coverage subject to orbital constraints) stands to benefit from both axes of \texttt{jaxsgp4}: differentiability for gradient-based optimisation and batch parallelism for evaluating candidate configurations at scale. 

\section*{Acknowledgements}
This work was supported by the UKRI GLITTER grant [EP/Y037332/1] and the UKRI Frontier Research Guarantee [EP/X035344/1]. The authors were supported by the research environment and infrastructure of the Handley Lab at the University of Cambridge.

\section*{Data Availability}

The source code for \texttt{jaxsgp4} is publicly available at \url{https://github.com/cmpriestley/jax_sgp4}. All benchmark scripts, data, and plotting code required to reproduce the figures in this paper are archived on Zenodo at \url{https://doi.org/10.5281/zenodo.19322209}.

\section*{Software and Tools}
The authors acknowledge the use of AI tools in preparing this manuscript and the accompanying software. Specifically, generative language models Gemini 3.1 Pro Preview (Google DeepMind) and Claude Opus 4.6 (Anthropic) were used to assist with drafting the text. The AI coding agent Claude Code (Anthropic) was used as a tool during software development. All AI-generated outputs were reviewed, edited, and validated by the authors to ensure their accuracy. All ideas, data, analysis and conclusions presented in this work are solely those of the authors.

\bibliographystyle{unsrt}  
\bibliography{references}

\end{document}